# Exciton spin relaxation in resonantly excited CdTe/ZnTe self-assembled quantum dots

S. Mackowski, T.A. Nguyen, T. Gurung, K. Hewaparkarama, H.E. Jackson, and L.M. Smith

*Department of Physics, University of Cincinnati, 24551-0011 Cincinnati OH*

J. Wrobel, K. Fronc, J. Kossut, and G. Karczewski

*Institute of Physics Polish Academy of Sciences, Warszawa, Poland*

## Abstract

*We study the exciton spin relaxation in CdTe self-assembled quantum dots (QDs) by using polarized photoluminescence (PL) spectroscopy in magnetic field. The experiments on single CdTe QDs and on large QD ensembles show that by combining LO phonon – assisted absorption with circularly polarized resonant excitation the spin-polarized excitons are photo-excited directly into the ground states of QDs. We find that for single symmetric QDs at B=0 T, where the exciton levels are degenerate, the spins randomize very rapidly, so that no net spin polarization is observed. In contrast, when this degeneracy is lifted by applying external magnetic field, optically created spin-polarized excitons maintain their polarization on a time scale much longer than the exciton recombination time. We also observe that the exciton spin polarization is conserved when the splitting between exciton states is caused by QD shape asymmetry. Similar behavior is found in a large ensemble of CdTe QDs. These results show that while exciton spins scatter rapidly between degenerate states, the spin relaxation time increases by orders of magnitude as the exciton spin states in a QD become non-degenerate. Finally, due to strong electronic confinement in CdTe QDs, the large spin polarization of excitons shows no dependence on the number of LO phonons emitted between the virtual state and the exciton ground state during the excitation.*

## INTRODUCTION

The ability to control and manipulate the spin of electrons in semiconductors is of paramount importance in the efforts to develop new methods and techniques for using the spin degree of freedom in electronic devices. This new field of research, called spintronics [1], anticipates large payoffs of new spin-based technology in information processing and data storage. In particular there is a strong interest in trying to use the spin degree of freedom in quantum dots (QDs) because of the significant suppression of spin relaxation processes expected for electrons localized to QDs [2]. In bulk semiconductors as well as semiconductor quantum wells, the spin relaxation times are found to be very short, typically of the order of picoseconds [3] due to elastic scattering of carriers (by phonons or impurities) that through momentum relaxation can effectively flip the spin [3]. Since carriers in QDs are confined in all three dimensions, the resulting full quantization of electronic states strongly suppresses such scattering processes. The resulting robustness of the exciton spin in QDs could then result in spin relaxation times of the order of several nanoseconds [4]. Recent time-resolved photoluminescence (PL) experiments performed on large QD ensembles under resonant excitation have shown that spin relaxation time of the excitons in QDs can indeed be as long as several nanoseconds: much longer than the exciton recombination time [5-8]. Importantly in this case, when the dots are populated resonantly, the presence of low-lying excited states or states within the wetting layer do not affect the exciton spin relaxation processes. Similar qualitative conclusions have been drawn from the non-resonant continuous-wave PL measurements of single QDs [9-12]. In this case, the long spin relaxation time has been inferred from the similar intensities of two single dot emission lines originating from exciton levels split by magnetic field. However, time-resolved PL experiments carried out on *single* asymmetric QDs have shown the spin relaxation time to be surprisingly short (~ 100 ps) [13]. This unexpectedly short spin relaxation time is probably caused by the quite weak spatial confinement of the "naturally formed" GaAs quantum dots.

These experimental efforts have been complemented by a number of theoretical studies that try to estimate the influence of different factors, such as interaction with nuclear spins, acoustic phonon scattering or spin-orbit coupling, on the spin confined in a QD [14-17]. For example, it has been found that as long as the spin-flip processes between Zeeman-split levels is considered, the dominant contribution comes from acoustic phonon emission between two exciton levels [17]. However, since phonons themselves do not flip the spins, the above process has to be mediated by spin-orbit interaction [16]. Similar mechanism determines the spin relaxation in the case of asymmetric QDs, where the exciton levels are split by the exchange interaction [14]; the energy range involved here (~ 1meV) is comparable to that of magnetic field induced Zeeman splitting of a QD-confined exciton.

From this work, it is clear that one of the requirement for long spin relaxation times of QD excitons, is that the excitons must be in the strong-confinement limit, where the excited state – ground state energy splitting is as large as possible. In this regard, ideally, the size of the QD should not be larger than the exciton Bohr radius. Moreover, it is also important to remove the closely coupled continuum of states in the wetting layer which exists in classical Stranski-Krastanov growth.

In this paper, we study exciton spin relaxation processes in CdTe QDs by means of resonantly excited PL spectroscopy. The optical experiments have been carried out for single CdTe QDs as well as for large QD ensembles. These self-assembled CdTe QDs are excellent candidates for exciton spin relaxation studies since there is no evidence for a wetting layer [18] and the dots are very small (~3 nm) in lateral size [19]. The absence of the continuum of states in a wetting layer eliminates the possibility of Auger-type spin scattering. In addition, the small size of QDs results in a very strong exciton confinement: recent PLE measurements of single CdTe QDs have shown that the average energy distance between excited and ground exciton states in these QDs is of the order of 100 meV [20]. Therefore, in order to study exciton spin relaxation in CdTe QDs, we utilize LO phonon-assisted absorption. Through circular polarization of the tunable laser with energy which is one to three LO phonon frequencies above the QD ground state we are able to *directly* excite spin-polarized excitons into the QDs.

The experimental results described in this paper show that the spin relaxation time of excitons in CdTe QDs is exquisitely sensitive to the degeneracy of the quantum-confined exciton spin states. In the case of degenerate states (symmetric QD at B=0 T) the QD emission is completely unpolarized regardless of the polarization of the excitation. This shows that exciton spin relaxation time is at least an order of magnitude less than the exciton recombination time. On the other hand, the application of even small magnetic fields or shape asymmetry of the QD potential lifts the degeneracy of exciton levels. In these cases the polarization of QD emission is predominantly the same as the polarization of the excitation. This means that the excitons exhibit spin relaxation times an order of magnitude larger than the exciton recombination time. We find an excellent agreement between the results obtained for single CdTe QDs and large QD ensembles. Through analysis of the polarized PL emitted from the QDs we find that the number of LO phonons involved in the excitation process does not influence the spin polarization of the QD –confined excitons.

## SAMPLES AND EXPERIMENT

The sample containing CdTe self-assembled QDs was grown by molecular beam epitaxy on a (100) oriented GaAs substrate. Four monolayers of CdTe were deposited by atomic layer epitaxy on a thick ZnTe buffer to form dots with a typical diameter of about 3 nm with a surface density of $10^{12}/cm^2$. Further details of the sample growth and characterization can be found elsewhere [18]. For the results discussed in this paper it is important to note that no two-dimensional uniform wetting layer has been detected for this sample using either power-dependent PL or PL excitation spectroscopy. Moreover, strong spatial confinement pushes the excited states towards very high energies, so one can use the LO phonon-assisted absorption to resonantly create excitons in these QDs.

Exciton spin polarization in CdTe QDs was examined by PL spectroscopy under resonant continuous-wave excitation by a Rhodamine 590 dye laser. The sample was placed on the cold finger of a continuous flow helium cryostat and maintained at a temperature of T=5 K. Magnetic fields up to B=4 T were applied in a Faraday configuration. Polarization of the excitation and emission was controlled by a combination of Glan-Thompson linear polarizers and Babinet-Soleil compensators. The emission was dispersed by a triple DILOR monochromator working in a subtractive mode and detected by a cooled CCD detector. The spectral resolution of the system is around 70 μeV. Resonant PL measurements were performed on large QD ensembles as well as on single CdTe QDs. Optical access to single dots was achieved by preparing small sub-micron apertures in an opaque metal mask and focusing the laser spot using a microscope objective. The excitation power in these measurements was low enough to assure that no more than one electron-hole pair is present in a QD at a time.

## EXPERIMENTAL RESULTS

In Fig. 1 we show resonantly excited PL obtained for a large ensemble of CdTe QDs at B=0 T (Fig. 1a,b) and B=4 T (Fig. 1c). These spectra show significant enhancement of the PL emission for those QDs whose ground states are one, two and three LO phonon energies below the energy of the excitation laser. Extensive PL excitation measurements on this sample have shown that the enhanced emission results from LO phonon-assisted absorption directly into these QDs ground states [20]. Therefore, with a suitable polarization of the excitation laser, we expect to excite spin-polarized excitons directly into these QDs. Therefore, through polarization analysis of the emitted PL, we can measure the spin polarization of the excitons at the time they recombine. Any reduction in the exciton spin polarization with respect to the polarization of the excitation must then result from spin relaxation processes that occur during the recombination time of the excitons.

In Fig. 1a**,** we show $\sigma^+$ and $\sigma^-$ polarized PL of CdTe QDs obtained at B=0 T with both $\sigma^+$ and $\sigma^-$ circularly polarized excitations. The integrated intensities as well as the overall shapes of $\sigma^+$ and $\sigma^-$ polarized emissions are the same for both circularly polarized excitations: the emitted PL is completely unpolarized. This indicates that for the circularly polarized excitation the exciton spins are completely randomized by the time they recombine, regardless of the initially excited spin distribution. Such a result requires that the time for the excitons to scatter between the degenerate states must be much shorter than the exciton recombination time which has been measured separately to be 300 ps [21].

Strikingly different behavior is seen in Fig. 1b when the excitation is linearly polarized along the [110] or [1−10] crystallographic directions ($\pi_X$ and $\pi_Y$, respectively [**9**]) and we analyze the linear polarization of the emission. In this configuration one can see that if the QDs are excited with $\pi_X$ - polarized laser, the QD emission is also predominantly $\pi_X$ - polarized. Similarly, when exciting with $\pi_Y$ – polarized laser, the emission is also $\pi_Y$ – polarized. On the other hand, when the excitation is circularly polarized (either $\sigma^+$ or $\sigma^-$) we observe no linear polarization anisotropy whatsoever; the intensities of $\pi_X$ and $\pi_Y$ – polarized emissions are identical.

In order to explain these results one has to consider two different sets of QDs present in the studied sample [18]. The first set consists of the dots that are cylindrically symmetric, so that the optically active exciton ground states (characterized by total angular momentum J=±1) are degenerate [9], as shown in the inset to Fig. 1a. On the other hand, the ground states of asymmetric (elongated) dots are symmetric and antisymmetric linear combinations of J=+1 and J=–1 states: X and Y (see the inset to Fig. 1b). Moreover, these two states X and Y are split by an exchange energy which depends on the degree of asymmetry of the dot [22]. Thus, for symmetric dots, the ground state excitons couple with $\sigma^+$ and $\sigma^-$ circularly polarized light, while the ground state excitons of asymmetric dots couple with $\pi_x$ and $\pi_y$ linearly polarized light. We note that when resonantly exciting a large ensemble of QDs through LO phonon-assisted absorption, both sets of QDs are populated. What set of QDs is probed is determined by the configuration of the excitation and emission polarizations. For example, when exciting with $\pi_x$-polarized light, there is equal probability of exciting J=+1 and J=–1 states in symmetric dots, but only the X states in the asymmetric dots are excited (see Fig. 1b). Correspondingly, with $\sigma$+ polarized light, there is equal probability of exciting X and Y states in asymmetric dots, but only J=+1 states are excited in symmetric dots (see Fig. 1a). We use this information to separately investigate the spin relaxation processes between these states through suitable polarization of the excitation laser and separately controlling the polarization of the emitted PL.

In an applied magnetic field the degenerate exciton levels (J=+1 and J=–1) of a symmetric QD are split by an energy proportional to the magnetic field: $\Delta E = g^{*}\mu_{B}B$ [9]. This splitting energy is determined by the effective g-factor for CdTe QDs which has been measured to be –3 [12]. Upon recombination the exciton states with J=±1 produce a right or left circularly polarized photons according to $\Delta J = \pm 1$ [9]. It is important to note, that the splitting does not influence the emission polarization of the exciton eigenstates in the symmetric QD [9]. In contrast, as discussed above, the exciton levels in the asymmetric QD are already split at zero magnetic field into the X and Y states. With applied magnetic field, this splitting increases due to Zeeman interaction and the polarization of the emission lines changes gradually from linear to circular [9]. The largest zero-field splitting observed for CdTe QDs studied here (~ 0.3 meV) are noticeably smaller than the magnetic field induced splitting at B=4 T (~ 0.6 meV). We expect then that at B=4 T the Zeeman splitting determines the selection rules of the emission so that almost all of the QDs in the ensemble would behave as symmetric ones. The comparison between PL results for single QDs and for the QD ensemble, which is discussed later, justifies this assumption.

In Fig. 1c we show resonantly excited PL spectra obtained for CdTe QDs at B=4 T. The excitation was either $\sigma^{+}$ (solid points) or $\sigma^{-}$ (open points) polarized and both circularly polarized components of the emission were analyzed. As can be seen, the polarization of the QD emission is predominantly the same as that of the excitation. This means that once the excitons are created in a single non-degenerate spin state, they essentially stay in that state before they recombine radiatively. This demonstrates that, similar to the case of asymmetric

QDs at B=0 T (see Fig. 1b), the spin relaxation time for these QDs at 4 T is much longer than the exciton recombination time (300 ps).

To summarize these results, we find that for degenerate exciton states (eg. the symmetric QD at B=0) the exciton spin relaxation time in CdTe QDs is much shorter than the recombination time. In contrast, when the exciton level degeneracy is lifted either by exchange interaction (asymmetric dots) or external magnetic field (symmetric dots at $B\neq0$) the exciton preserves its spin polarization throughout its lifetime. Since the recombination time of excitons in these QDs is equal to 300 ps, we suppose the spin relaxation times in these two cases to be of the order of 10 ps to 1 ns, respectively. From these results we also conclude that the asymmetric QDs in the ensemble are preferentially aligned along either [110] or [1-10] crystallographic directions [22].

Looking more closely at the data in Fig. 1c it appears that the response of the QDs is different for $\sigma^+$ or $\sigma^-$ excitation. In particular, the degree of polarization of the emitted PL for $\sigma^+$ excitation is less than for $\sigma^-$ excitation. This asymmetry is clearly due to the partial thermalization of the excitons to the lower energy ($\sigma^-$ polarized) state. It is important to note that the net transition rate from higher energy spin state to lower energy spin state is slightly larger than the reverse process by exactly the Boltzmann factor exp $(-\Delta E/k_BT)$ where $\Delta E$ is the energy splitting between the spin states in magnetic field. Even though thermal equilibrium is clearly never achieved, the spins cool down (from $\sigma^+$ to $\sigma^-$ polarized state) more efficiently than they warm up (from $\sigma^-$ to $\sigma^+$ polarized state) [23].

In order to study in more detail the effect of using LO phonon-assisted absorption to pump spin-polarized excitons into the QDs, we performed PL measurements under resonant circularly polarized excitation on *single* CdTe QDs using fixed nano-apertures as discussed previously. Studying single QDs simplifies the analysis, since we remove the possible influence of the size, chemical composition or symmetry variations within the whole QD ensemble on the measured polarization characteristics. In Fig. 2 we show the nano-PL spectrum excited at E=2.115 eV at B=0 T obtained for a 0.5 μm diameter aperture. Both $\sigma^+$ and $\sigma^-$ circular polarizations for excitation and detection were used, similarly as for the macro-PL studies discussed previously. Apart from three broad LO phonon resonances, a large number of narrow lines corresponding to exciton recombination in single CdTe QDs is seen. In agreement with the PL results on large QD ensembles, essentially no preferential circular polarization of the emission is observed at B=0 T for the symmetric single QD, as shown in the inset to Fig. 2.

In Fig. 3 we present the nano-PL spectra measured in the same way for three different single symmetric CdTe QDs at B =2.5 T. These three QDs (marked in Fig. 2 by the shaded areas) are carefully chosen to have the emission energy at B=0T approximately 1-, 2- and 3-LO phonons (Fig. 3a-c, respectively) away from the excitation laser energy at 2.115 eV. We point out that the spectra for all three single QDs presented in Fig. 3 were taken with the same excitation energy. All three QDs show strong PL intensity for identical circular polarizations of excitation and detection. In contrast, when the circular polarizations of excitation and detection are opposite, the emission intensity is significantly weaker. This finding supports the results described previously: that in applied magnetic field the spin relaxation time of the excitons confined in CdTe QDs is much longer than the exciton recombination time. It is also important to note, that the degree of spin polarization observed for these QDs is approximately the same, indicating that the spin relaxation processes are *independent* on the excitation energy.

To determine that these QDs are *not* excited through an excited state but through LO phonon assisted absorption, we adjusted the dye laser slightly to insure that the intensity of the emission lines is not strongly sensitive to the excitation energy. Previous nano-PLE measurements show that the linewidths of excited state resonances are around 200-600 μeV (comparable to the Zeeman splitting) whereas the LO phonon-assisted absorption line is nearly an order of magnitude broader (see Fig. 2) [20]. This enables us, by controlling the polarization of the excitation laser, to precisely select which one of the Zeeman-split exciton spin state is populated.

These single-dot measurements clearly illuminate the results obtained for the whole QD ensemble. At zero magnetic field, when the exciton levels in a symmetric QD are degenerate, the exciton spins completely randomize within the exciton lifetime, indicating extremely short spin relaxation times. In external magnetic field, as expected, the single dot emission line splits, according to effective exciton g-factor of –3. With the spin degeneracy removed in this way, we find that the exciton spin relaxation is strongly suppressed.

**DISCUSSION**

In order to analyze the suppression of spin relaxation in QDs with increasing magnetic field, we fit spectra similar to those shown in Fig. 1 with four Gaussian lines: three of them representing the LO phonon replicas and the fourth representing the relatively weak background emission. The excitation spectrum of this background emission, as evidenced by detailed PL studies, reflects the shape (energy and linewidth) of non-resonantly excited PL [20]. We interpret this background PL as resulting from direct excited - state ground state excitations in QDs. Consequently, we fit the resonantly excited spectra assuming that both the energy and the linewidth of this emission are the same as those of non-resonantly excited

PL [18]. Furthermore, the fits were performed under an additional assumption that this background emission is unpolarized regardless of the polarization of the excitation. This is not too surprising, as a particular excitation energy should match only randomly a given excited state. Furthermore, we have no detailed information about the magnetic field influence on excited states in our QDs, in particular, the value and the sign of the splitting, as well as polarization selection rules. In the absence of any information, we therefore assume the emission associated with direct excited state – ground state excitation is unpolarized, however it does not appear that the results described next are sensitive to this assumption. Of course, the results of resonant PL obtained for single CdTe QDs are obviously unaffected by any of these assumptions since the background is not present.

Using this fitting procedure, the intensities of all three LO phonon replicas for all four polarization configurations are obtained as a function of magnetic field. We define the polarization as $P=(I^+ - I^-)/(I^+ + I^-)$, where $I^+$ and $I^-$ correspond to intensity of the first, second or third LO phonon replica in $\sigma^+$ and $\sigma^-$ polarizations, respectively. In Fig. 4 we show the absolute values of the polarization, $|P|$, as a function of magnetic field for all three phonon replicas measured for both $\sigma^+$ and $\sigma^-$ polarized excitations. With increasing magnetic field, $|P|$ shows a strong increase and tends to saturate at about 70% when approaching B=4 T. These results clearly indicate that the exciton spin relaxation is a strong function of the degeneracy of the exciton spin states in QDs. When both exciton spin states are degenerate (symmetric QD at B=0 T), the spin relaxation time must be much less than the exciton recombination time so that the initially polarized exciton spin distributions completely randomize before recombination. On the other hand, when this degeneracy is lifted the spin relaxation time increases by orders of magnitude and becomes much larger than the exciton recombination time.

Importantly, the value of |P| does not depend on the number of LO phonons involved in the QD ground state excitation. The insensitivity of the exciton spin polarization to the number of LO phonons involved in the excitation could indicate that any spin relaxation observed in the experiment occurs after populating the QDs ground states. Similar insensitivity to LO phonons is also observed in the linearly polarized excitation and detection (see Fig. 1b) experiments. This is quite different from the case of semiconductor quantum wells as well as in bulk semiconductors where it has been found that multiple LO phonons scatterings may contribute to the depolarization of excitons [24,25]. However in these experiments the relaxation by LO phonons occurs through *real* electronic states, with phonons necessary for momentum conservation. In contrast, for the CdTe QDs described here, the phonon cascade does not involve any real electronic states apart from the final ground state. In addition, due to strong spatial confinement, there is no dispersion of momentum for excitons in these QDs.

While no dependence of the polarization on the number of LO phonons involved in the excitation is observed for the fixed circular polarization of the excitation, the emission excited with $\sigma^-$ polarization is somewhat more strongly polarized than the other one. The slight preference for $\sigma^-$ excitation reflects the fact that it is always the $\sigma^-$ – polarized states, which are lower in energy, so that the relaxation rate is from the state higher in energy to the lower energy state is slightly larger than for the reverse process. Cooling of spins seems to be then more effective than heating [23].

Recently similar experiments at zero field were reported by Scheibner *et al.* [8] on CdSe self-assembled QDs. In those experiments, a linearly polarized excitation was used to excite directly the asymmetric dots at zero field. The experiments were performed on relatively shallow dots, and they found that the polarization of the emission depended strongly on the number of phonons emitted in the excitation. The emission of QDs featured 50% polarization for the first LO phonon replica, and the degree of polarization appeared to decrease by 20% with each subsequent emission of a phonon. However, the calculation of polarization reported there assumed that the entire PL results from LO phonon-assisted absorption, neglecting the possibility of resonant excitation of the QDs through excited state-ground state transition. These CdSe dots were also significantly larger and shallower than the CdTe QDs discussed here, so that the excited states are expected to be lower in energy. In this regard, it is important to note that our strongest evidence that the spin polarization of excitons in QDs is independent of the number of phonons involved in the QD excitation is our direct measurement of the polarization of *single* CdTe QDs. As shown in Fig. 3, in magnetic field polarization of the single symmetric QD emission is nearly complete, regardless whether one- two- or three- LO phonon emissions are involved. Importantly, when probing excitons in single QDs one can neglect the influence of any additional component of the emission. We add the calculated polarization values obtained for single QDs as stars in Fig. 4. The agreement between these two datasets (nano- and macro-PL measurements) is excellent above around B=2T. At lower magnetic field the values of P obtained for the whole ensemble measurements are slightly smaller than those obtained for single CdTe QDs. We ascribe this difference to asymmetric QDs, since at moderate magnetic fields the emission characteristic for asymmetric QDs is not completely circularly polarized [9]. However, as discussed previously, at high magnetic field, when the exchange splitting due to asymmetry becomes smaller than the Zeeman splitting, the emission is almost completely circularly polarized (as for symmetric QDs). This explains the agreement of polarization values at high magnetic field for both single QDs and the large QD ensemble. We conclude that the results of resonant spectroscopy of both QD ensemble and single CdTe QDs show that the spin polarization of excitons in QDs does not depend on the number of LO phonons participating in the excitation process.

**CONCLUSIONS**

In conclusion, using polarized magneto-photoluminescence spectroscopy we study the low temperature spin relaxation of excitons in CdTe QDs as a function of magnetic field. When the exciton spin states in QDs are degenerate, the spin relaxation time is much shorter than the exciton recombination time, so that no net polarization of QD emission is observed. As soon as the exciton level degeneracy is lifted, however, either by symmetry or external magnetic field, the exciton spin levels become decoupled, with the spin relaxation time much longer than the recombination time of the exciton. In this case the emission of QDs is polarized identically to the excitation. The polarization of QD emission increases rapidly with increasing magnetic field. We also find that, when taking into account influence of direct excitation through excited states, the spin polarization of excitons in QDs is not affected by the number of LO phonons participating in the excitation process. The results of resonant polarized PL obtained for the whole QD ensemble are in almost perfect consistency with those of single CdTe QDs.

**ACKNOWLEDGEMENT**

The work was supported by NSF grants nr 9975655 and 0071797 (United States). Partial support through grant PBZ-KBN-044/P03/2001 and project SPINOSA (Poland) is acknowledged.

**Figure captions**

**FIG. 1.** Resonantly excited macro-PL spectra of CdTe QDs: (a) circularly polarized excitation and detection at B=0T, (b) linearly polarized excitation and detection at B=0T, (c) circularly polarized excitation and detection at B=4T. Schematic energy diagrams of excitons in QDs for each experimental configuration are displayed as insets.

**FIG. 2.** Nano-PL of CdTe QDs measured at B=0 T. The shaded areas show emission lines from QDs for which magnetic field dependencies are presented later. In the inset, polarized spectra at zero field are displayed for the 2.064 eV QD emission line which was excited through 2-LO phonon absorption.

**FIG. 3.** Magnetic field nano-PL at 2.5 T obtained with circularly polarized excitation and detection for three single QDs marked in Fig. 2. These dots are populated through LO phonon-assisted absorption with the emission of one, two and three LO phonons.

**FIG. 4.** Polarization, P, of the CdTe QDs emission plotted as a function of magnetic field. The data obtained from macro-PL and nano-PL (stars) is shown. For the former the dependencies measured for the first (squares), second (circles) and third (diamonds) LO phonon replicas are presented. Values obtained for $\sigma^+$ ($\sigma^-$) – polarized excitation are represented by solid (open) symbols.

S. Mackowski *et al.*, Figure 1

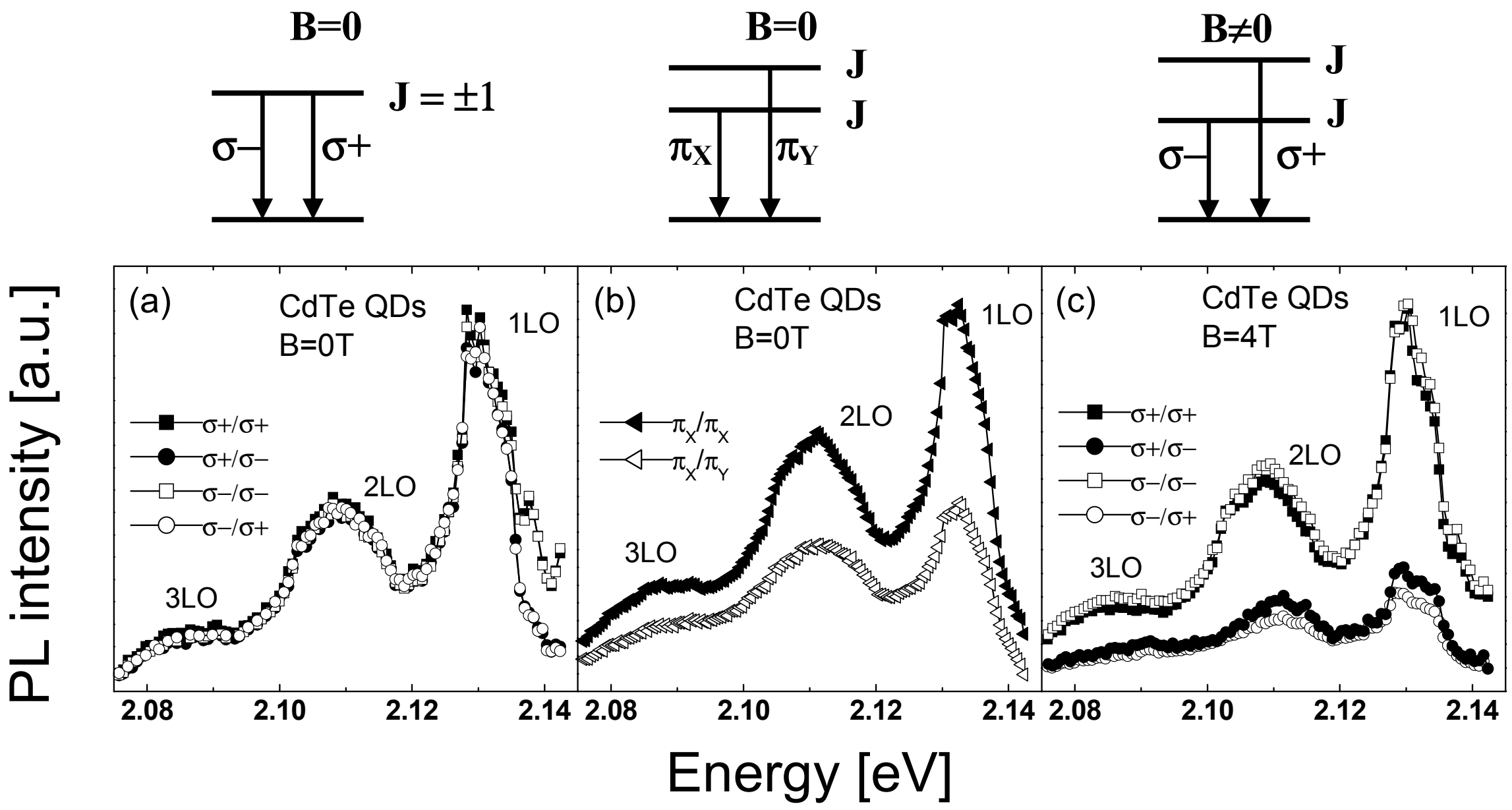

S. Mackowski *et al.*, Figure 2

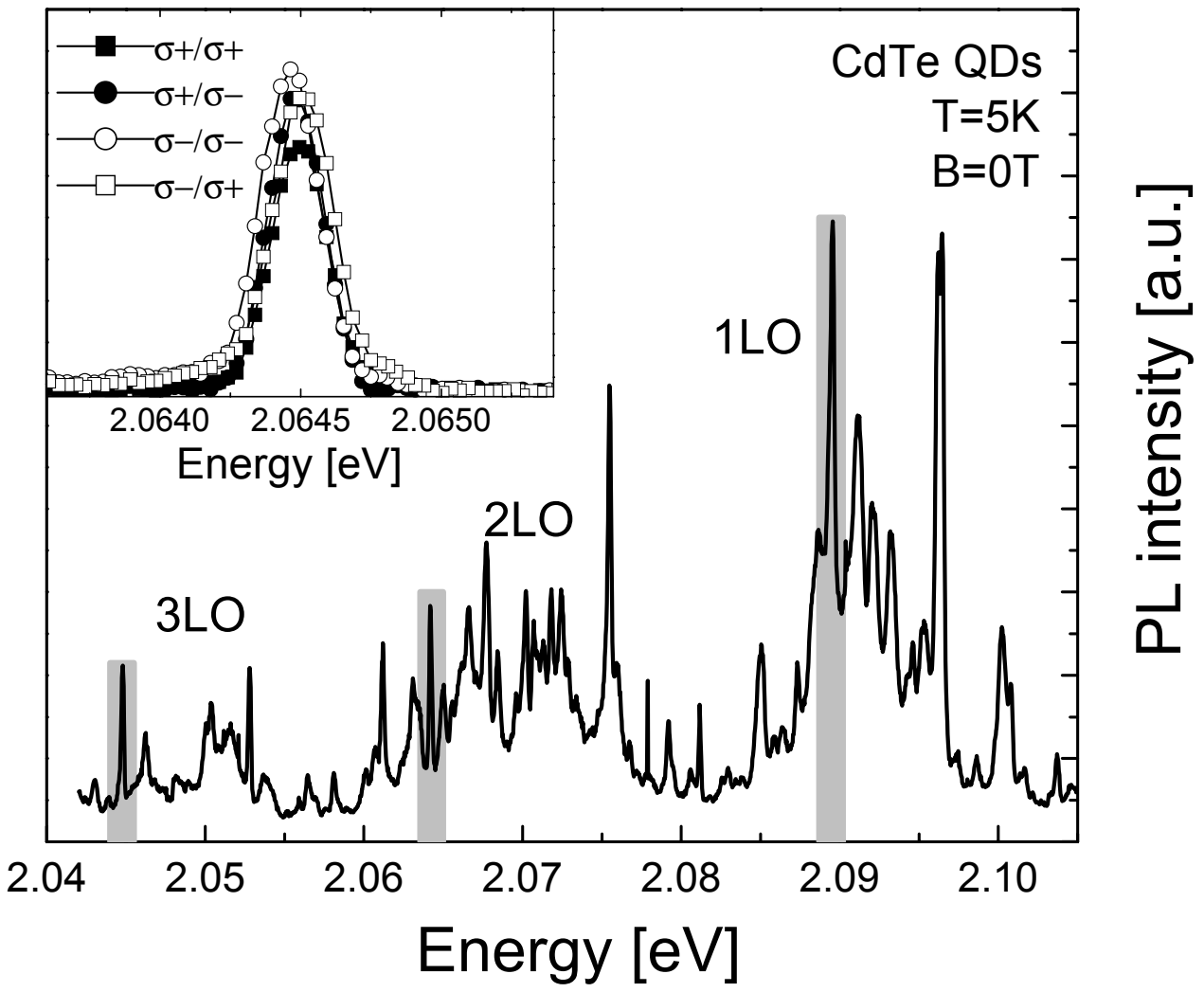

S. Mackowski *et al.*, Figure 3

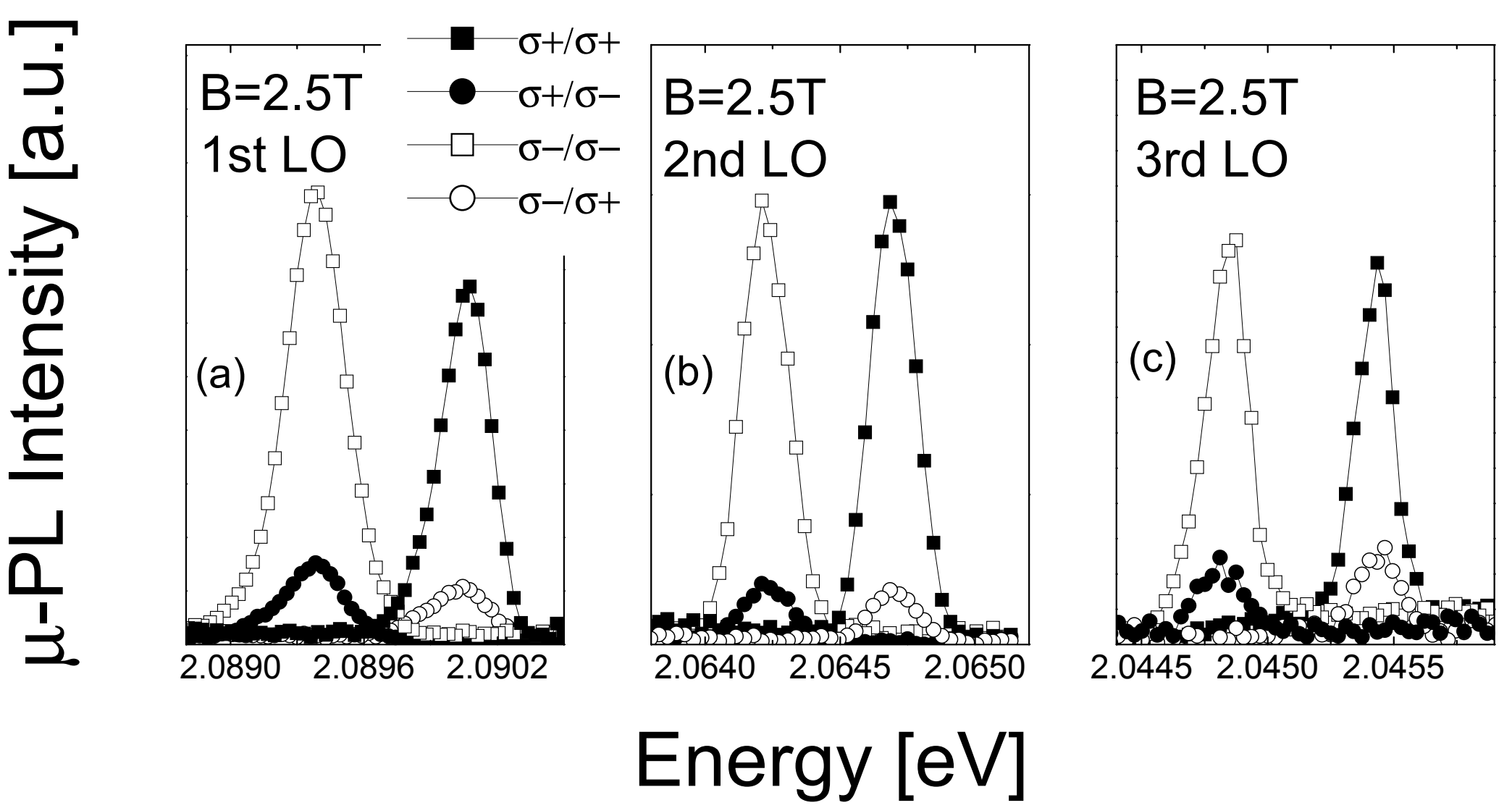

S. Mackowski *et al.*, Figure 4

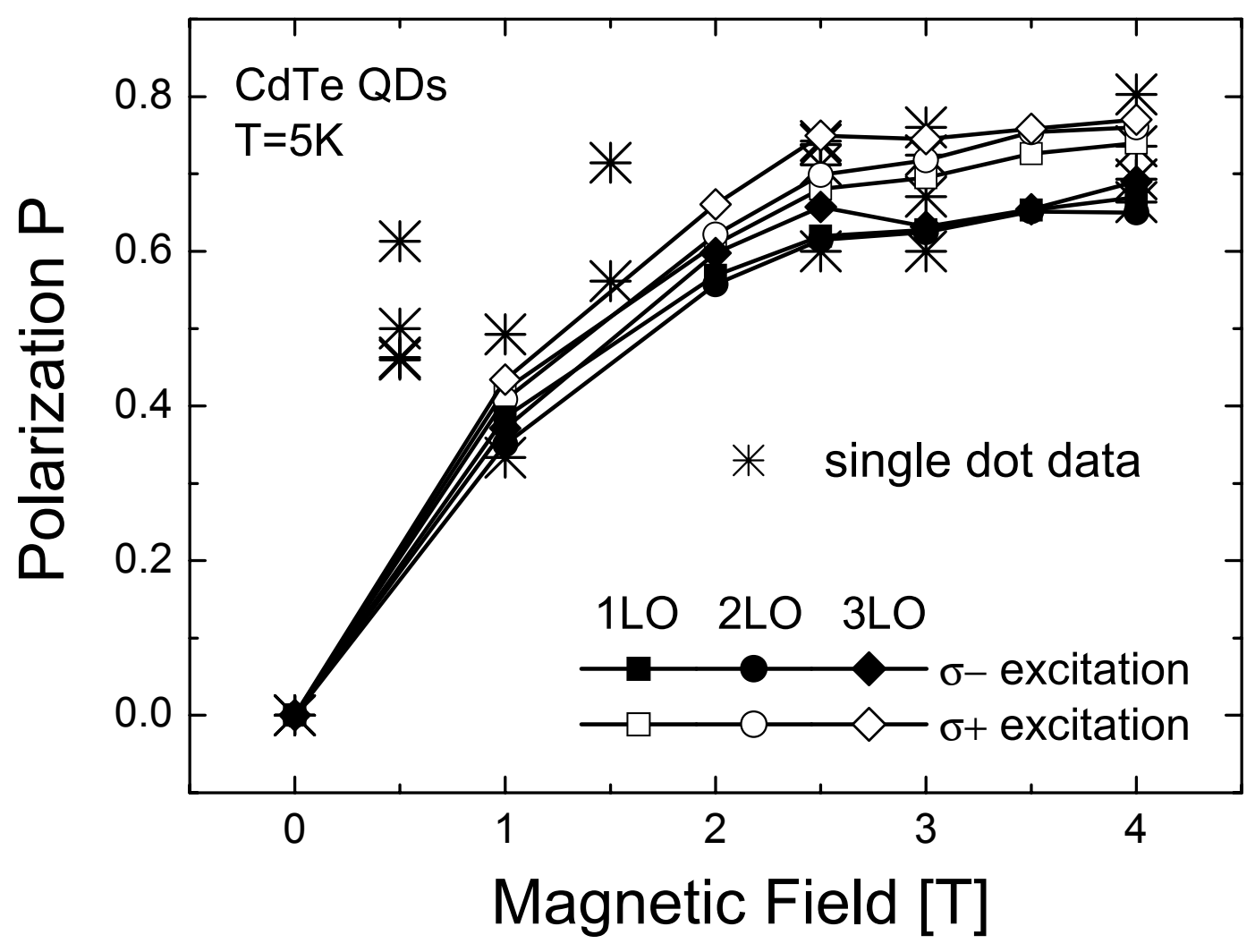